\documentstyle[aps,preprint,prl]{revtex}
\begin{document}
\flushbottom

\widetext
\draft
\title{Suppression of Heavy Ion $\gamma \gamma$ Production of the Higgs
by Coulomb Dissociation}
\author{A. J. Baltz}
\address{
Physics Department,
Brookhaven National Laboratory,
Upton, New York 11973}
\author{Mark Strikman}
\address{
Department of Physics,
Pennsylvania State University,
University Park, Pennsylvania 16802}
\date{\today}
\maketitle

\def\thepage{\arabic{page}}
\makeatletter
\global\@specialpagefalse
\ifnum\c@page=1
\def\@oddhead{Draft\hfill To be submitted to Phys. Rev. C}
\else
%\ifnum\c@page>1
\def\@oddhead{\hfill}
\fi
\let\@evenhead\@oddhead
\def\@oddfoot{\reset@font\rm\hfill \thepage \hfill}
\let\@evenfoot\@oddfoot
\makeatother

\begin{abstract}
Predicted two-photon Higgs production with heavy ions at LHC is shown to be
reduced due to the large Coulomb dissociation cross
section.  Incorporating the effect of dissociation reduces the production of
a 100 GeV Higgs by about a factor of three compared to rates in the
literature calculated without this effect.
\end{abstract}
%\newpage

\makeatletter
\global\@specialpagefalse
\def\@oddhead{\hfill}
\let\@evenhead\@oddhead
\makeatother
\nopagebreak
%\twocolumn
%\narrowtext
The possible production of the Higgs particle or other heavy particles via the 
coherent two-photon mechanism from colliding heavy ion beams at LHC has been a
subject of much interest in recent
years\cite{dre,pap89,gra,bau,pap90,cah,nor,mul}.
However, with the exception of one recent work,\cite{henck} the
modification of production rates due to Coulomb dissociation of the nucleus
has been ignored.  Henken, Trautmann, and Baur\cite{henck} calculated the
effective
$\gamma \gamma$ luminosity in conjunction with giant dipole excitation of one
of the nuclei and found this higher order process appreciable when compared to
the $\gamma \gamma$ luminosity calculated without consideration of other
processes.  In this note we investigate the effective suppression
of Higgs production at LHC due to interference of Coulomb
dissociation not only via the giant dipole state but also through
equivalent photons of up to many GeV impinging on each nucleus in its rest
frame\cite{brw}.  The large magnitude of these higher excitations is seen in
the recent calculated cross sections for Coulomb dissociation in
Pb + Pb collisions at LHC: including all excitations led to 220 barns; 
including only the giant dipole excitation led to 127 barns\cite{brw}.

In the standard calculation the two
colliding heavy ions (e. g. Pb + Pb) are assumed to travel on straight line
trajectories at an impact parameter such that their densities do not overlap.
Each of the ions produce a spectrum (equivalent photon number) of
Weizsacker-Williams photons of energy $\omega$ dependent
on the transverse distance $b_i$
\begin{equation}
N(\omega,b_i)={  Z^2 \alpha \omega^2 \over \pi^2 \gamma^2 }
K_1^2({b_i\,\omega \over  \gamma})
\end{equation}
where $K_1$ is the modified Bessel function and $\gamma$ is the relativistic
factor of the colliding ions seen in the center of mass frame.
The effective $\gamma \gamma$ luminosity function at a given equivalent mass
$W$ is then given by\cite{bau,pap90}
\begin{equation}
L_{\gamma \gamma}(W) = 2 \pi \int {d \omega_1 \over  \omega_1 }
\int_{R_1}^\infty b_1\, d b_1 
\int_{R_2}^\infty b_2\, d b_2 
\int_0^{2 \pi} d \phi\, N_1 (\omega_1,b_1)\, N_2 ({W^2 \over 4 \omega_1},b_2)
\,\theta(b - R_1 -R_2)
\end{equation}
where $R_1$ and $R_2$ are the nuclear radii and $b$ is the ion-ion impact
parameter
\begin{equation}
b^2 = b_1^2 + b_2^2 - 2 b_1 b_2 \cos(\phi).
\end{equation}
The $\theta$ function excludes impact parameters where densities overlap.
The cross section for producing a particle in the heavy ion
collision is then
\begin{equation}
\sigma(W) = {8 \pi^2 \over W^3}\, \Gamma_{H \rightarrow \gamma \gamma}(W)\,
L_{\gamma \gamma}(W) 
\end{equation}
where $\Gamma_{H \rightarrow \gamma \gamma}(W)$ is the two photon decay width
of the Higgs.
 
From Fig. 2 of
Ref.\cite{brw} one can see that the probability of a colliding Pb ion being
dissociated is in the field of the other Pb ion at LHC is approximately equal
to $[1. - exp(-(17.4/b)^2)]$ where $b$, the impact parameter, is in fermis.
The survival probability (neither ion being Coulomb dissociated) is then
approximately $exp(-2(17.4/b)^2)$.  A parallel calculation including only
the giant dipole resonance gives a corresponding survival probability of
approximately $exp(-2(11.2/b)^2)$. 

Figure 1 shows the effect of Coulomb dissociation on the luminosity function
for the 
%$gamma 
$\gamma
= 3000$ of LHC.  $R_1$ and $R_2$ were set at 7 fm.
The upper curve is the luminosity without dissociation,
the middle curve shows the luminosity reduced by Coulomb dissociation
via the giant dipole resonance, and the lower curve includes Coulomb
dissociation to all final states.

We now calculate Higgs production at LHC.  Calculation of the width
$\Gamma_{H \rightarrow \gamma \gamma}(W)$ is a textbook
exercise\cite{ell,bar,daw}.  The
mechanism is dominated by triangle loops of which the W$^{\pm}$ is most
dominant
followed by the top quark.  Lower mass contributions are relatively
insignificant and we have ignored them here.  Figure 2 shows the effect of
Coulomb dissociation on Higgs production.  The cusp at 160 GeV is at 
twice the mass of the W${^\pm}$.  At 100 GeV the production rate
of the 
%higgs 
Higgs
is reduced by more than a factor of three from the rate
calculated without Coulomb dissociation. 
%%%
Note that the effective suppression factor depends on the kind of detector
used to select the $\gamma \gamma$ mechanism.  If one uses the lack of
activity in the 
zero angle calorimeter the suppression factor is as we calculated above.
On the other hand if one uses a detector with a wide rapidity covarage
like one discussed for the FELIX detector\cite{FELIX} the
Coulomb dissociation would lead to much less of a suppression.

Note also that the 
calculated rates are fairly
sensitive to the radius and impact parameter cutoff.  If we set
$R_1$ and $R_2$ to 8 fm rather than 7, then the 100 GeV Higgs is reduced
by 41\% on the top curve and by 30\% on the bottom curve.
Such an increase in radius is maybe justified by a large  ($\sim 2$)
increase of the radius of the 
strong interaction at LHC energies\cite{abe} as compared to the incident
energies $\sim$1 GeV which were used to determine the effective
nuclear radii for $pA$ interactions.

One of us (AJB) would like to acknowledge useful conversations with Sally
Dawson.

\vskip .5cm
This manuscript has been authored under Contract No. DE-AC02-76-CH00016 with
the U. S. Department of Energy. 
The work was partially supported
 by Department of Energy under Contract No. DE-FG02-93ER40771.

\begin{figure}
\caption[Figure 1]{$\gamma \gamma$ luminosity function.
The upper curve is without dissociation,
the middle curve includes Coulomb dissociation only
via the giant dipole resonance, and the lower curve includes Coulomb
dissociation to all final states.}
\label{higg}
\end{figure}
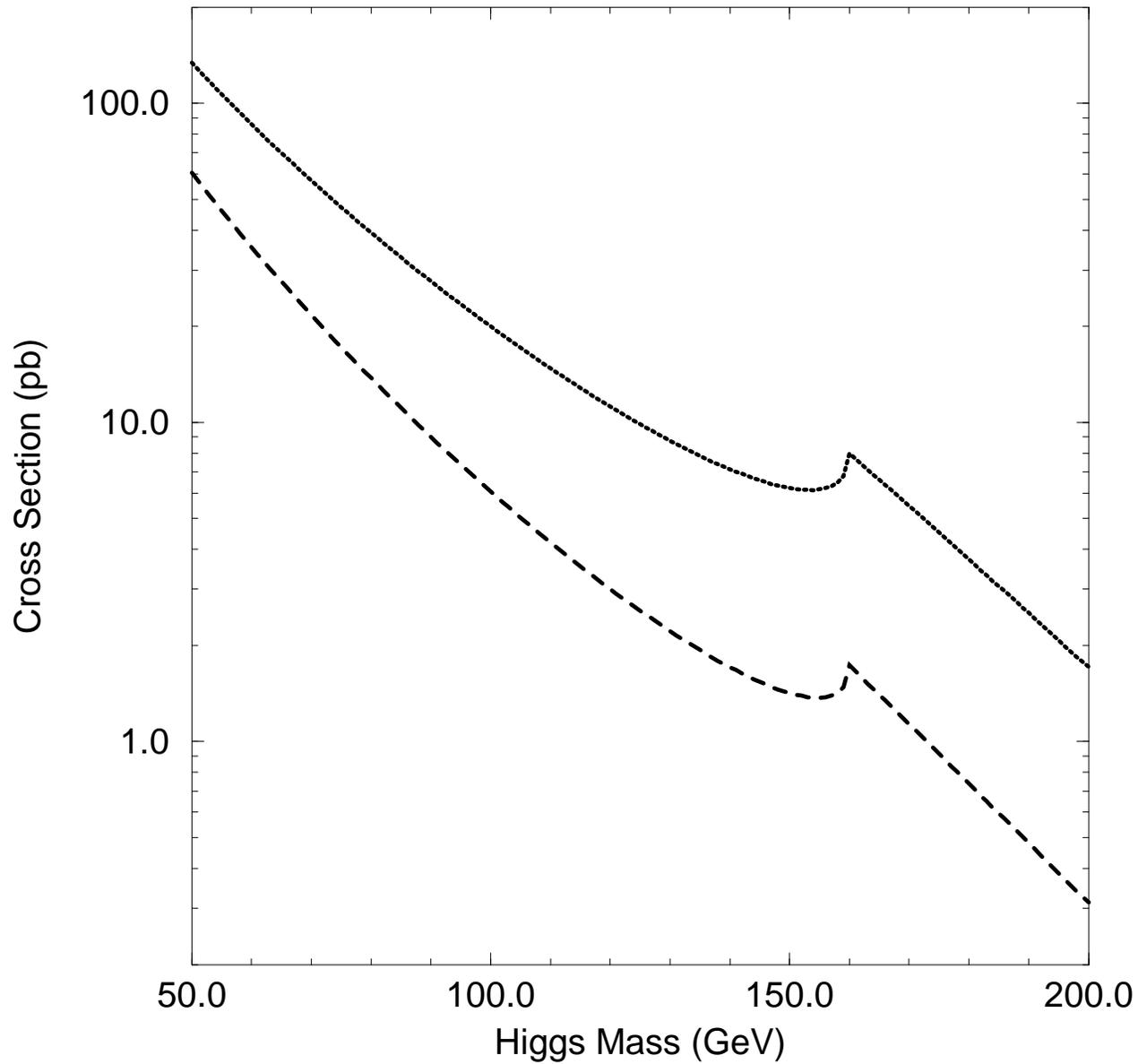
\begin{figure}
\caption[Figure 2]{Coherent Electromagnetic Higgs Production at LHC.
The upper curve is without dissociation, and the lower curve includes Coulomb
dissociation to all final states.}
\label{lumi}
\end{figure}
\end{document}